\documentstyle[prb,aps]{revtex}
\begin{document}
\draft
\tighten

\title{Recursion and Path-Integral Approaches to the Analytic Study of the
Electronic Properties of $C_{60}$}

\author{Yeong-Lieh Lin and Franco Nori}
\address{Department of Physics, The University of Michigan, Ann Arbor, MI
48109-1120}

\maketitle
\begin{abstract}
The recursion and path-integral methods are applied to analytically
study the electronic structure of a neutral $C_{60}$ molecule.  We
employ a tight-binding Hamiltonian which considers both the $s$ and $p$
valence electrons of carbon. From the recursion method, we obtain
closed-form {\it analytic} expressions for the $\pi$ and $\sigma$
eigenvalues and eigenfunctions, including the highest occupied
molecular orbital (HOMO) and the lowest unoccupied molecular orbital
(LUMO) states, and the Green's functions. We also present the local
densities of states around several ring clusters, which can be probed
experimentally by using, for instance, a scanning tunneling microscope.
{}From a path-integral method, identical results for the energy spectrum
are also derived. In addition, the local density of states on one
carbon atom is obtained; from this we can derive the degree of
degeneracy of the energy levels.
\end{abstract}

\pacs{PACS numbers: 02.90.+p, 31.15.+q}

\widetext
\section{Introduction}
Since the recent discovery of a simple technique for the production in
bulk quantities of fullerenes, undoped and doped $C_{60}$ molecules
have generated enormous interest among chemists, physicists and
materials scientists. In $C_{60}$, carbon atoms sit at the sixty
vertices of the pentagons and hexagons of a truncated-icosahedron. The
carbon-carbon bonds are of two different lengths: $1.46$ \AA \  for the
single bonds (bonds on the pentagons) and $1.40$ \AA \ for the double
bonds (bonds on the hexagons not shared by a pentagon). The single
(double) bonds are denoted by solid (dotted) lines throughout the
figures in this paper. The literature on $C_{60}$ is vast, and here we
do not attempt a review. The interested reader is referred to the
references \cite{1,2,3,4,5,6,7} and papers listed therein.

Most studies on the electronic structure of $C_{60}$ have been mainly
numerical. In the present work, we focus on an analytical study of the
electronic properties of a single neutral $C_{60}$ molecule. We model
this system through a tight-binding Hamiltonian (Eq.~(1)) which
considers both the $s$ and $p$ valence electrons of carbon. This paper
is organized as follows.  In Sec.~II, we discuss the physical nature
and types of the interaction between these valence electrons, with a
total number of $240$. The couplings can be separated into two main
contributions: coming from the $\pi$-bonded and $\sigma$-bonded
electrons.

We first focus on the $\pi$ states. In Sec.~III we use the recursion
method\cite{8} to analytically solve the Hamiltonian for the $\pi$
electrons. We derive {\it closed-form expressions} for their energy
eigenvaluess and eigenfunctions. The beauty of the recursion method for
$C_{60}$ lies in the fact that the recurrence terminates very quickly
({\em e.g.}, after four iterations only), providing exact and very
concise expressions for the parameters of the recursion. The same
results for the $\pi$-states energy spectrum are derived in Sec.~IV by
using a path-integral method. This approach follows Feynman's
programme: to compute physical quantities from sums over paths. In
addition, the local density of states (LDOS) on one carbon atom is
obtained; from this we can derive the degree of degeneracy of the
energy levels.

In the two sections mentioned above, every step of each approach is
exact. All the calculations are done analytically, either by hand or
with the assistance of computer symbolic manipulation software.
Diagonalizations are fulfilled by iteratively applying the Hamiltonian
on initial states. The relations between the recursion and
path-integral methods are discussed in Sec.~V. It can be seen that
while these two approaches are related to each other, each one has its
own advantages. These  methods are significantly different from the
ones used so far. Furthermore, they are neither numerical nor require
the use of group theory. For a very lucid and clear account of a
different approach, based on group theory and focused only on the $\pi$
states, the reader is referred to Ref.~[3].

In Sec.~VI, we present the precise algebraic expressions as well as
figures for the LDOS around several ring clusters ({\em i.e.}, a
carbon atom, a pentagon ring, and a hexagon ring; also for two
opposite carbon atoms, pentagons, and hexagons). We find that around
a pentagon (hexagon) ring the LDOS is large at low (high) energies.
This is related to the fact that a pentagon (hexagon) has zero
(three) double bonds and five (three) single bonds. These
quantities are relevant to the several important experimental
techniques which probe the local spectroscopy of molecules;
(for instance, by using a scanning tunneling microscope, as
described in the review in Ref.~[9]). In Sec.~VII we discuss
the solutions of the $\sigma$-states Hamiltonian Eq.~(3). To
describe the $\sigma$ states, the coupling between orbitals
on the same atom and that between orbitals along the same bond
are both taken into account. By following a ``one band---two band"
transformation\cite{10}, the eigenvalues and eigenfunctions
for the $\sigma$ states are then analytically obtained.

\section{Electronic states}

To investigate the electronic properties of a single $C_{60}$ molecule, we
consider the four carbon valence electrons $2s$, $2p_{x}$, $2p_{y}$, and
$2p_{z}$ in the tight-binding Hamiltonian
\begin{equation}
{\bf H}_{T}\ =\ \sum_{i,\alpha} \ \epsilon_{\alpha}\ c_{i\alpha}^{+}
c_{i\alpha}\ +\ \sum_{<ij>,\alpha,\beta} \ t_{\alpha\beta}\
c_{i\alpha}^{+}c_{j\beta}.
\end{equation}
Here, $i$ denotes the carbon site and $\alpha$ denotes the valence
orbitals $2s$, $2p_{x}$, $2p_{y}$, and $2p_{z}$. Also,
$\epsilon_{\alpha}$ is the orbital energy and $t_{\alpha\beta}$ is the
hopping matrix element between orbitals on the nearest-neighboring
sites $i$ and $j$. The sixty $2p_{z}$ orbitals, each pointing along the
outward radial direction, are hybridized to form $\pi$ states. The
other three orbitals $2s$, $2p_{x}$ and $2p_{y}$, distributed on the
plane tangential to the surface of the molecule, are hybridized along
the lattice bonds to form $\sigma$ bonding and antibonding states. Let
us assume that the mixture of these three orbitals at each site produce
three $sp^{2}$ hybrid orbitals: $sp^{2}_{a}$ along the double bond and
$sp^{2}_{b}$  and $sp^{2}_{c}$ along the two single bonds,
respectively. From a physical point of view, the $60$ outer $\pi$
orbitals are relevant to the conducting properties of the molecules and
the $180$ $\sigma$ orbitals are mainly responsible for the elastic
properties. The former are also responsible for the bond dimerization.
Also, only $\pi$-states occur around the Fermi energy.

As a result of its planar structure, the $\pi$ electrons in graphite
contain only pure $2p_{z}$ orbitals. Nevertheless in the curved
structure of $C_{60}$, an extra small component from $2s$, $2p_{x}$ and
$2p_{y}$ is induced along the radial direction due to the surface
curvature. The interaction between the $\pi$ orbitals in $C_{60}$ are
then slightly increased in comparison to that in graphite. However,
since this component is very small, the overlap integral between $\pi$
and $\sigma$ orbitals is negligible. The nonzero Hamiltonian parameters
used here are the same ones as in Ref.~[2]. Therefore, the original
Hamiltonian (1) can be written in terms of ${\bf H}_{\pi}$ and ${\bf
H}_{\sigma}$ as
${\bf H}_{T}\ =\ {\bf H}_{\pi}\ \oplus\ {\bf H}_{\sigma},$
where
\begin{equation}
{\bf H}_{\pi}=- \sum_{<ij>}\ t_{ij}\ c_{i}^{+}\,c_{j},
\end{equation}
and
\begin{equation}
{\bf H}_{\sigma}=-V_{1}\ \sum_{i,\alpha \neq \beta} \
c_{i\alpha}^{+}c_{i\beta}\ -\ V_{2}\ \sum_{<ij>} \  c_{i\alpha}^{+}c_{j\alpha}.
\end{equation}
The electronic $\pi$-states Hamiltonian (2) describes the kinetic energy of the
$2p_{z}$ electrons hopping on the sixty vertices of a $C_{60}$ fullerene and
$t_{ij}$ is the hopping integral between nearest-neighboring atoms
$i$ and $j$. The hopping integrals are considered unequal for single and double
bonds. Also $c_{i}^{+}$ is the electron operator creating a $2p_{z}$ orbital on
the atom located at vertex $i$. A constant term corresponding to the on-site
$2p_{z}$ orbital energy is omitted in Eq.~(2). In the $\sigma$-states
Hamiltonian Eq.~(3), $\alpha=sp^{2}_{a}, sp^{2}_{b}$ and $sp^{2}_{c}$ denotes
the hybridized orbitals. Also, $V_{1}$ stands for the hopping integral between
orbitals on the same carbon atom and $V_{2}$ stands for that between orbitals
on different atoms that are  associated with the same bond (the length
difference between single and double bonds is neglected here). A more detailed
discussion on the origin of $V_{1}$ and $V_{2}$ is presented in Appendix A.

Admittedly, this is a  simplified model (like the ``Ising model") for the
electronic structure of $C_{60}$, the ``Hydrogen atom" of the main
fullerene family of $C_{60{\cdot}n^{2}}$ molecules. However, its
understanding is a convenient stepping stone to the study of more complex
models. The spectroscopy of, for instance, the Hydrogen atom can be
solved by using several different analytical approaches. These lead
to the same analytical expressions for the eigenvalues and eigenvectors. In
spite of the fact that modern computers can easily obtain numerical
expressions for them, it is useful to have analytic results. In fact,
analytical solutions have always been pursued in the area of
spectroscopy of atoms, molecules, and clusters. In this paper,
we pursue several approaches to the analytical study of the
spectroscopy of an important molecule. We would like to
emphasize that the approaches described here are not intended
to substitute more traditional methods, but to present
alternative viewpoints and complementary results.

\section{Recursion method approach}

\subsection{Formulation}
We first focus on the $\pi$ states Hamiltonian (2). Let us begin
with a brief outline of the recursion method \cite{8} for obtaining the
eigenvalues
and eigenfunctions. First, one must choose an appropriate
normalized starting
state $|f_{0}>$. Further states are iteratively generated by the
three-term recurrence relation
\begin{equation}
{\bf H}\,|f_{n}>=a_{n}\,|f_{n}>+b_{n+1}\,|f_{n+1}>+b_{n}\,|f_{n-1}>,
\end{equation}
with the condition $b_{0}|f_{-1}>\equiv0$. Here, the real
parameters $a_{n}$ and $b_{n+1}$ are
determined by
\begin{equation}
a_{n}=<f_{n}|{\bf H}|f_{n}>
\end{equation} and
\begin{equation}
b_{n+1}=<f_{n}|{\bf H}|f_{n+1}>=<f_{n+1}|{\bf H}|f_{n}>.
\end{equation}
By convention, the $b$'s are chosen to be positive.
The process terminates at $|u_{N-1}>$ with the last recurrence
${\bf H}|f_{N-1}>=a_{N-1}|f_{N-1}>+b_{N-1}|f_{N-2}>.$
The constructed orthonormal states
\{$|f_{0}>,|f_{1}>,\ldots,|f_{N-1}>$\}
along with the parameters
\{$a_{0},a_{1},\ldots,a_{N-1}$\} and
\{$b_{1},b_{2},\ldots,b_{N-1}$\} thus constitute
the ``chain model" \cite{8} of a given Hamiltonian.
The representation of ${\bf H}$ in the new basis
$\{|f_{n}>\}$ is in a tridiagonal form.

The energy levels can be achieved by constructing the following
polynomials with $P_{-1}(E)=0$ and $P_{0}(E)=1$,
\begin{equation}
P_{n+1}(E)=\frac{(E-a_{n})P_{n}(E)-b_{n}P_{n-1}(E)}{b_{n+1}}.
\end{equation}
The eigenvalues $E_{\lambda}$ are determined by the $N$
zeros of the last polynomial $P_{N}(E)=0$, with an arbitrary
nonzero $b_{N}$. It follows then that the eigenfunctions are
\begin{equation}
\frac{1}{\cal{N}_{\lambda}}\sum_{n=0}^{N-1} P_{n}(E_{\lambda})|f_{n}>,
\end{equation}
where
${\cal N}_{\lambda}=[\sum_{n=0}^{N-1} P_{n}^{2}(E_{\lambda})]^{1/2}$
are the normalization constants for the  eigenfunctions, and
$P_{0}(E_{\lambda})$ always equals $1$.

So far, we have briefly outlined the recursion method.
We now come to the application of this approach to the electronic
structure of a fullerene molecule. For convenience, we work in
units of the single-bond hopping-integral. The hopping amplitude
is therefore $1$ for every single bond and $t$ for each double
bond. It has been pointed out \cite{1} that the value of $t$ is
about $1.1$. We will use two alternative sets
(denoted by $\cal A$ and $\cal B$) of starting states. Each
set consists of two initial states from which the whole energy
spectrum can be derived and the final results do not depend
on the choice of initial bases. Both sets lead to the
{\it same} solutions for
the eigenvalues and eigenfunctions. From the physical point
of view, this consistency ensures the equivalent interpretation
of the results from these two sets of initial states.
We denote by $|j>$ the $2p_{z}$ orbital centered at the $j$th-atom.
For convenience of visualization, in all our figures we flatten
the truncated icosahedral structure of $C_{60}$ into a plane.
Note that labellings vary according to the different choices
of initial states.

\subsection{Case $\cal A$}

The first set of initial states
starts from a $5$-atom pentagon ring (see Fig.~1(a)) and
a $6$-atom hexagon ring (see Fig.~1(b)).
Starting from a pentagon ring, we choose the initial state
$|u_{0}>$ as a linear combination of the five orbitals on it
\begin{equation}
|u_{0}>=\frac{1}{\sqrt{5}}\,\sum_{j=1}^{5}\ |j>.
\end{equation}
Here the labelling is as shown in Fig.~1(a). From now on,
${\bf H}$ will donote ${\bf H}_{\pi}$ unless specified
otherwise ({\em e.g.}, ${\bf H}_{\sigma}$). As an illustration,
we present here the calculation of the first recurrence.
\begin{eqnarray*}
{\bf H}\,|u_{0}>& = &-\frac{1}{\sqrt{5}}\,[(|2>+|5>+t\,|6>)+
(|1>+|3>+t\,|7>) \\
&   & \mbox{}+(|2>+|4>+t\,|8>)+(|3>+|5>+t\,|9>)+(|1>+|4>+t\,|10>)] \\
&  = & -2\ [\frac{1}{\sqrt{5}}\,(|1>+|2>+|3>+|4>+|5>)]+t\
[-\frac{1}{\sqrt{5}}\,(|6>+|7>+|8>+|9>+|10>)] \\
&  =& -2\ |u_{0}>+t\ |u_{1}>.
\end{eqnarray*}
We then obtain $a_{0}=-2$, $b_{1}=t$ and
$|u_{1}>=-\frac{1}{\sqrt{5}}\sum_{j=6}^{10}\ |j>.$
Following the same procedure, we can construct further
states and obtain the parameters $a$'s and $b$'s at the same time.
The beauty of the recursion method for $C_{60}$ lies in
the fact that the recurrence terminates very quickly,
exactly at $|u_{7}>$.
As a result, we have exact and very concise formulas for $a_{0}$
through $a_{7}$, and $b_{1}$ through $b_{7}$.
Similarly, starting from a linear combination
of the six orbitals on a hexagon ring, we have the initial
state $|v_{0}>$ as
\begin{equation}
|v_{0}>=\frac{1}{\sqrt{6}}\,\sum_{j=1}^{6}\ (-1)^{j+1}|j>,
\end{equation}
where the labelling is referred to Fig.~1(b). The choice of
alternating signs, $(-1)^{j+1}$, meets the three-fold symmetry
requirement of a hexagon ring in $C_{60}$, with alternating single
and double bonds. Again, the recurrence terminates at $|v_{7}>$.
We then obtain
another group of parameters $a$'s and $b$'s. All these $|u_{n}>$
and $|v_{n}>$ states, as well as their respective parameters $a_{n}$
and $b_{n+1}$, are summarized in Table I.

Through these two groups of $a$'s and $b$'s, we can
respectively construct
two polynomials $P_{8}^{p}(E)$ and $P_{8}^{h}(E)$
which are of eighth degree in $E$. Here the superscript $p$,
for pentagon, ($h$ for hexagon) refers to the polynomial
constructed from the group of $a$'s and $b$'s
generated by $|u_{0}>$
($|v_{0}>$). The roots of these two polynomials
can be analytically obtained and correspond
to the electronic energy levels. It follows then that we have a
total of $16$ distinct eigenvalues.
When $t=1$, a common root $-1$ exists for both polynomials
$P_{8}^{p}$ and $P_{8}^{h}$. We thus have $15$ distinct energy levels.
It is also straightforward to
obtain the eigenvectors through Eq.~(6).

\subsection{Case $\cal B$}

In this approach, we exploit the
symmetry property that the inversion operator leaves the
$C_{60}$ molecule invariant. We therefore take the first
(second) starting state $|\phi_{0}>$ ($|\varphi_{0}>$) as
a linear combination of the orbitals
on two opposite ({\it i.e.}, antipodes) pentagon (hexagon) rings
\begin{equation}
|\phi_{0}>=\frac{1}{\sqrt{10}}\,\sum_{j=1}^{5}\ [|j>+ {\cal P} |j'>]
\end{equation} and
\begin{equation}
|\varphi_{0}>=\frac{1}{\sqrt{12}}\,
\sum_{j=1}^{6}\ (-1)^{j+1}\,[|j>+ {\cal P} |j'>],
\end{equation}
where the site labels are presented in Fig.~1(a) (1(b)) for the first
(second) starting state. Here $\cal P$ stands for parity, with
the value $+1$ or $-1$.
Note that atom $j$ and atom $j'$
are antipodes.
Following the same procedure, we find the very convenient
and remarkable result that {\em the recurrence terminates even faster} at
$|\phi_{3}>$ ($|\varphi_{3}>$) for the first (second)
initial state. Due to the two possible values of
$\cal P$, we have two different results for
$a_{3}$ for each starting state. In Appendix B we present the calculation
of the last recurrence, namely those for $|\phi_{3}>$ and $|\varphi_{3}>$.
It can be clearly seen that ${\bf H}\,|\phi_{3}>$ generates two possible
values for $a_{3}$ and no further state; and so does
${\bf H}\,|\varphi_{3}>$. As a consequence, we
obtain four groups of \{$a_{0},a_{1},a_{2},a_{3}$\} and
\{$b_{1},b_{2},b_{3}$\}. All the states $|\phi_{n}>$ and
$|\varphi_{n}>$, as well as their
respective parameters $a_{n}$ and $b_{n+1}$ are listed
in Table II. It follows that we can construct four
polynomials $P_{4}^{p+}(E)$, $P_{4}^{p-}(E)$,
$P_{4}^{h+}(E)$ and $P_{4}^{h-}(E)$. Each one is
fourth degree in $E$. The superscript $p+$ stands
for the polynomial constructed from  $|\phi_{0}>$
with $\cal P$ equal to $+1$, and similarly for the others.
With the choice of $b_{4}=1/b_{1}b_{2}b_{3}$,
these polynomials can be explicitly written as
\widetext
$$P_{4}^{p+}(E)=E^{4}+(2\,t+3)\,E^{3}+(5\,t-1)\,E^{2}-
(2\,t^{3}-t^{2}+8)\,E-(t+2)(t^{3}-t^{2}+t+2),$$
$$P_{4}^{p-}(E)=E^{4}+3\,E^{3}-(2\,t^{2}-t+1)\,E^{2}-
(3\,t^{2}-4t+8)\,E+(t^{4}-t^{3}+t^{2}+4\,t-4),$$
$$P_{4}^{h+}(E)=E^{4}-2\,(t+1)\,E^{3}+(3\,t-1)\,E^{2}-
(2\,t^{3}+t+2)\,E-(t^{2}+1)(t^{2}+t-1),$$
and
$$P_{4}^{h-}(E)=E^{4}-2E^{3}-(2t^{2}+t+1)E^{2}-
(2t^{2}+t+2)E-(t+1)^{2}(t^{2}-t+1).$$

\noindent
By analytically solving these four polynomials, we obtain the
same $16$ eigenvalues obtained above (case $\cal A$).
Similarly we can obtain the eigenvectors, which are also
equal to those obtained from the alternative $\cal A$.

\subsection{Results}

In Table III, we summarize
the eigenvalues and the corresponding eigenvectors
for the case $t=1$. In Table IV, we present
the closed-form eigenvalues explicitly expressed in
terms of the single-bond hopping integral $t_{1}$
and the double-bond hopping integral $t_{2}$. Thus, eigenvalues for the
limiting cases $t_{1}=0$ or $t_{2}=0$ can also be readily
inferred. As to the eigenfunctions, we present those for
the highest occupied molecular orbital (HOMO) and the
lowest unoccupied molecular orbital (LUMO) in terms of $t$, the ratio of the
hopping matrix elements for single and double bonds.
Since the respective degree of degeneracy for each
eigenvalue can be acquired from the local density of states on a carbon
atom (discussed in a later section), the other degenerate
eigenvectors can be generated by standard group
theory analysis. This is outside the scope of the present paper. We
therefore only present these eigenfunctions derived from the pure
application of the recursion method. It is instructive to notice that
the following points further support the results for the degeneracy.
They are: (1) the sum of the product of energy and its corresponding
egeneracy equals the trace of the Hamiltonian, which is zero; (2) the number
of states with even parity equals that with odd parity, and (3) the
behavior of the eigenvalues can be easily studied in the limits when either
$t_{1}=0$ or $t_{2}=0$.

\subsection{Relations between alternatives $\cal A$ and $\cal B$}

It is worthwhile to point out the following
relations between the cases $\cal A$
and $\cal B$ presented before. First,
$P_{8}^{p}(E)=P_{4}^{p+}(E)P_{4}^{p-}(E)$
and  $P_{8}^{h}(E)=P_{4}^{h+}(E)P_{4}^{h-}(E)$
up to an overall constant factor.
In other words, roots solved from $P_{8}^{p}(E)$
($P_{8}^{h}(E)$) are identical to those solved from
$P_{4}^{p+}(E)$ and $P_{4}^{p-}(E)$
($P_{4}^{h+}(E)$ and $P_{4}^{h-}(E)$).
Second, from Tables I and II we can see that, for $n=0$, $1$, $2$ and $3$,
$$|\phi_{n}>=\frac{1}{\sqrt{2}}\,(|u_{n}>-{\cal P}|u_{7-n}>),$$
and
$$|\varphi_{n}>=\frac{1}{\sqrt{2}}\,(|v_{n}>+{\cal P}|v_{7-n}>).$$
Third, for those eigenvalues $E_{\lambda}$ which are
common roots of $P_{8}^{p}$ and $P_{4}^{p+}$,
or common roots of $P_{8}^{h}$ and $P_{4}^{h-}$,
$$P_{7-n}(E_{\lambda})=-P_{n}(E_{\lambda}).$$
Also for those eigenvalues $E_{\lambda}$ which are
common roots of $P_{8}^{p}$ and $P_{4}^{p-}$,
or common roots of $P_{8}^{h}$ and $P_{4}^{h+}$,
$$P_{7-n}(E_{\lambda})=P_{n}(E_{\lambda}).$$
Here, all the $P_{n}(E_{\lambda})$'s
refer to the polynomials constructed in
the case $\cal A$ and $n=0, 1, 2$ and $3$.
Fourth, $P_{1}(E_{\lambda})$, $P_{2}(E_{\lambda})$ and
$P_{3}(E_{\lambda})$ calculated from the two alternatives
$\cal A$ and $\cal B$ are the same.
Fifth, for the eigenvector with respect to the
same eigenvalue, the normalization constants $\cal N_{\cal A}$
and $\cal N_{\cal B}$, calculated in
$\cal A$ and $\cal B$ respectively,
satisfy ${\cal N}_{{\cal A}}=\sqrt{2}\,{\cal N}_{{\cal B}}$.
{}From the
above five properties, the equivalence of results from
both alternatives becomes clear.
As an example, for an eigenvalue
$E_{\lambda}$ which is a common root of
$P_{8}^{p}$ and $P_{4}^{p+}$, we can obtain
the eigenvector $|\Psi_{\lambda}>$from $\cal A$ as
\widetext
\begin{eqnarray*}
 |\Psi_{\lambda}>& = &\frac{1}{{\cal N_{\cal A}}}\,\sum_{n=0}^{7}\
P_{n}(E_{\lambda})|u_{n}> \\
&  = & \frac{\sqrt{2}}{{\cal N_{{\cal A}}}}
\ [P_{0}\,\frac{1}{\sqrt{2}}\,(|u_{0}>-|u_{7}>)+
P_{1}\,\frac{1}{\sqrt{2}}\,(|u_{1}>-|u_{6}>) \\
& & \mbox+P_{2}\,\frac{1}{\sqrt{2}}\,(|u_{2}>-|u_{5}>)+
P_{3}\,\frac{1}{\sqrt{2}}\,(|u_{3}>-|u_{4}>)] \\
&  = &\frac{1}{{\cal N_{\cal B}}}\,\sum_{n=0}^{3}\
P_{n}(E_{\lambda})|\phi_{n}>,
\end{eqnarray*}

\noindent
where the last equality is just the result
obtained directly from $\cal B$.
In fact, from the results
of alternative $\cal A$, we can understand the parity
property associated with $C_{60}$.

\section{Path-integral (or moment) approach}

\subsection{Methodology and application}

In this method, the first and central task is the computation
of path-integrals (moments), defined by
\begin{equation}
{\cal M}_{l}\equiv <j|{\bf H}^{l}|j>,
\end{equation}
where the order $l$ is a positive integer. Note that ${\cal M}_{0}=1$
when $l=0$. The physical meaning of the above quantum-mechanical
expectation
value is as follows. The Hamiltonian ${\bf H}$ is applied
$l$ times to an initial $2p_{z}$ electron state $|j>$, localized
at carbon site $j$. Each time ${\bf H}$ is applied, the
electron gains a certain amount of kinetic energy depending
upon the bond (single or double) it travels. This
enables the electron to hop through $l$ bonds, reaching the
final state ${\bf H}^{l}|j>$. The path-integral (moment) just equals the
total kinetic energy gained by the electron returning to
the starting site $j$ after hopping $l$ steps. It is
obvious that ${\cal M}_{l}$ will be zero when the $l$-hops path
does not return to the starting site. In other words, ${\cal M}_{l}=0$ when
there
is no path of $l$ hops for which the electron may return
to the initial site. For the case $t=1$, the absolute value of
the moment ${\cal M}_{l}$ is the total number of closed paths
of $l$-steps starting and ending at the same site. This approach
follows Feynman's programme: to compute physical quantities
from sums over paths.

The path-integrals can be calculated analytically by hand,
as well as by computer using a symbolic manipulation program.
Below we describe both ways. Let us first examine the action
of the Hamiltonian on an arbitrary state $|j>$. This results
in three nearest-neighbor atom states with an additional
factor accounting for the respective bond hopping energy.
For example, ${\bf H}\,|1>=(-t)\,|2>+(-1)\,|3>+(-1)\,|3''>$
(labels as shown in Fig.~2). For simplicity and without any
loss of generality, we choose orbital $|1>$ to be
our starting state. Now, for $l=1$, starting from vertex
$1$ and following the connectivity of $C_{60}$ (Fig.~2),
we write down the factors $-t$, $-1$ and $-1$ on the vertices
$2$, $3$ and $3''$ respectively. A similar procedure holds for $l=2$.
Starting from the three resulting vertices
with respective factors, we then write down $(-t)(-1)=t$
on the vertices $4$, $4'$, $4''$ and $4'''$; $(-1)(-1)=1$
on the vertices $5$ and $5''$; and
$(-1)(-1)+(-1)(-1)+(-t)(-t)=2+t^{2}$ on the vertex $1$.
Our strategy is: each time the power of the Hamiltonian
({\it i.e.} kinetic energy) increases by one, we
move to the adjacent nearest-neighboring vertices. Also
the factor on each vertex is the sum of
the factors on the three nearest-neighbor
vertices times the bond hopping integral between vertices.
It is then straightforward, with the aid of ``the flattened
$C_{60}$ graph", to obtain all the factors on the
available vertices for any power of the Hamiltonian.
{}From its definition, it is evident that the path-integral of
order $l$ is just the factor on the vertex $1$ for
${\bf H}^{l}$. For example, ${\cal M}_{1}=0$
({\it i.e.}, the number of closed paths obtained by moving
one step is zero) and
${\cal M}_{2}=2+t^{2}$. By following this strategy,
we can generate, one by one, all the path-integrals to any order.
It is worthwhile to notice the mirror-symmetry between
the left and right halves of $C_{60}$ for the
atom vertices and bonds. So we need only concern ourselves with the
factors on the vertices in the right part. Furthermore,
because of the geometrical equivalence, with respect to
vertex $1$, of eight pairs of vertices ($j$
and $j'$ in Fig.~2), the total number of independent
vertices can be reduced to $24$. All calculations up to this point
can be done analytically by hand.

An alternative analogous procedure can be implemented by using symbolic
manipulation software on a computer. First, we define an auxiliary
quantity \cite{11}, $W_{l}(j)$, which is the sum over all possible
paths of $l$ steps on which an electron may hop from the
vertex $1$ to the vertex $j$. From the connectivity
of $C_{60}$, we can then construct $24$ independent
recurrence relations. For instance, $W_{l+1}(1)=-t\,W_{l}(2)-2\,W_{l}(3)$,
$W_{l+1}(2)=-t\,W_{l}(1)-2\,W_{l}(4)$, and
$W_{l+1}(3)=-t\,W_{l}(4)-W_{l}(1)-W_{l}(5)$. The recurrence
relations state that the vertex $j$ can be reached by taking the
($l+1$)th step from the three
nearest-neighbor vertices. The factors $-t$ and $-1$
account for the connecting bond-hopping integral. With
the initial conditions $W_{0}(1)=1$ and $W_{0}(j)=0$
for the rest of $j$'s, we can obtain the path-integrals to any order as
${\cal M}_{l}=W_{l}(1)$. We list ${\cal M}_{3}$
through ${\cal M}_{8}$ here: ~$0$, ~$t^{4}+8\,t^{2}+6$ $(15)$,~
$-2$,~ $t^{6}+18\,t^{4}+4\,t^{3}+48\,t^{2}+20$ $(91)$,
{}~$-14\,t^{2}-14$ $(-28)$,
{}~$t^{8}+32\,t^{6}+16\,t^{5}+184\,t^{4}+48\,t^{3}+256\,t^{2}+70$ $(607)$, for
$l=3,\cdots,8$. The numbers in parentheses are the values
when $t=1$. The correctness of the calculated path-integrals is
assured by the consistency of the results from these two
approaches. It is evident that through these two approaches we
can also obtain the quantities $<1|{\bf H}^{l}|j>$ for
$j \neq 1$ which can be appropriately interpreted as
the ``{\it sum-over-paths}" between sites $1$ and $j$.
For instance, $<1|{\bf H}^{l}|j>$ just equals $W_{l}(j)$.

To obtain the energy spectrum, we again utilize equation~(7). So the
next step is to express the parameters $a_{n}$ and $b_{n+1}$ in
terms of the moments. We employ the following formulas \cite{12}:
define the auxiliary matrix $M$ with the
first row elements defined as
$M_{0,l}\equiv{\cal M}_{l}$. The other rows are evaluated
by using only one immediate predecessor row:
\widetext
\begin{equation}
M_{n,l}=\frac{M_{n-1,l+2}-M_{n-1,1}\,M_{n-1,l+1}}
{M_{n-1,2}-M_{n-1,1}^{2}}-
\sum_{k=0}^{l-1}\ M_{n,k}\,M_{n-1,l-k}~,~~~n\geq 1;l=0,1,\cdots.
\end{equation}

\noindent
The $a_{n}$'s and $b_{n+1}$'s are obtained from the elements
of the second and third columns as
\begin{equation}
a_{n}=M_{n,1}
\end{equation}
and
\begin{equation}
b_{n+1}^{2}=M_{n,2}-M_{n,1}^{2},
\end{equation}
here $n=0,1,2,\cdots.$  Note that elements in the first column
$M_{n,0}$ are always equal to 1.
We analytically find that $b_{15}^{2}$ {\em exactly}
equals $0$ for $t=1$ and $b_{16}^{2}$
{\em exactly} equals $0$ for an arbitrary $t$.
Below we discuss the $t=1$ case. Similar conclusions can
be drawn for $t\neq 1$. The appearance of absolute zero
for $b_{15}^{2}$ in the case $t=1$ indicates the truncation at
$b_{15}$. In this sense, we expect $15$ eigenvalues. It also turns
out that the highest order of moment we need is $l=30$.
The moments ${\cal M}_{9}$ through ${\cal M}_{30}$ for $t=1$
are presented in Table VI.
Through the calculated parameters
\{$a_{0},a_{1},\ldots,a_{14}$\} and
\{$b_{1},b_{2},\ldots,b_{14}$\} (listed in Table VII),
we can construct the
polynomial $P_{15}(E)$ by using Eq.~(5). By solving $P_{15}(E)=0$,
we thus obtain $15$ energy levels. The results are
exactly identical to those obtained from the
recursion method approaches.
We present ${\cal M}_{9}$ through ${\cal M}_{16}$ for
an arbitrary $t$ in Table VIII.

\subsection{Alternative application of the path-integral method}

In the above description of the path-integral approach, the main
ingredient is
the computation of the path-integrals for a natural choice of state
$|1>$ centered at the
atom labelled by $1$. However, it is worthwhile to incorporate the
inversion
symmetry property. Therefore, instead of focusing on a single
localized
state, we turn to the computation of the path-integrals
with respect to states
\begin{equation}
 |I_{\pm}>=\frac{1}{\sqrt{2}}\,(|1> \pm |24>),
\end{equation}
where atoms $1$ and $24$ are antipodes (as labelled in Fig.~2).
It is a simple exercise to construct the following identity
for the path-integrals defined by $<I_{\pm}|{\bf H}^{l}|I_{\pm}>$,
\begin{equation}
 <I_{\pm}|{\bf H}^{l}|I_{\pm}>=<1|{\bf H}^{l}|1> \pm <1|{\bf H}^{l}|24>.
\end{equation}
These moments can then be readily obtained since the quantities
$<1|{\bf H}^{l}|1>$ and $<1|{\bf H}^{l}|24>$ are already
available. Note that the lowest order for the appearance of
a nonzero $<1|{\bf H}^{l}|24>$ is $l=9$. This is because the shortest path
for vertex $1$ to reach vertex $24$ contains $9$ steps.
We thus obviously have
$<I_{\pm}|{\bf H}^{l}|I_{\pm}>={\cal M}_{l}$ for $l \leq 8$.

For the case $t=1$, the highest order we need for
$<I_{+}|{\bf H}^{l}|I_{+}>$ ($<I_{-}|{\bf H}^{l}|I_{-}>$) is $l=14$
($16$), because $b_{7}^{2}$ ($b_{8}^{2}$) calculated from moments
$<I_{+}|{\bf H}^{l}|I_{+}>$ ($<I_{-}|{\bf H}^{l}|I_{-}>$) turns
out to be exactly $0$. Through the computed parameters
\{$a_{0},\ldots,a_{6}$\} and \{$b_{1},\ldots,b_{6}$\}
(\{$a_{0},\ldots,a_{7}$\} and \{$b_{1},\ldots,b_{7}$\}),
we {\it analytically} obtain $7$ ($8$) eigenvalues which are
identical to those belonging to the ${\cal P}=+1$ (${\cal P}=-1$) category
from the recursion method. Moments $<I_{+}|{\bf H}^{l}|I_{+}>$
($l=9,\cdots,14$) and $<I_{-}|{\bf H}^{l}|I_{-}>$ ($l=9,\cdots,16$)
are presented in Table IX and their respective set of parameters $a_{n}$'s
and $b_{n+1}$'s in Table X. The result that $a_{n}$ and $b_{n+1}$ for
$n=0,1,2$ and $3$ in Table X are identical to those in Table VII comes
from the fact $<I_{\pm}|{\bf H}^{l}|I_{\pm}>={\cal M}_{l}$ for
$l \leq 8$. For an arbitrary $t$, we need moments
$<I_{\pm}|{\bf H}^{l}|I_{\pm}>$ up to order $16$ (
$<I_{\pm}|{\bf H}^{9}|I_{\pm}>$ through
$<I_{\pm}|{\bf H}^{16}|I_{\pm}>$ are listed in Table XI).
We analytically find that $b_{8}^{2}$ calculated from these two
sets of moments {\it exactly} equals $0$ in both cases.
Consequently, consistent results for the eigenvalues are recovered.
In this section, we have presented an unconventional choice of
initial states and concentrated on the moments with respect to
these states. It is shown that this approach is even more efficient
in analytically obtaining the energy eigenvalues.

\section{Relationship between the recursion and path-integral methods}

Generally speaking, the path-integral method is closely related
to the recursion method, especially in the aspect that both
methods lead to the same results
for the parameters $a_{n}$ and $b_{n+1}^{2}$.
In this section, we illuminate this point by showing that the same
expressions for the parameters in the recursion method
case ${\cal B}$ can be obtained through the moment method. As
the initial states are $|\phi_{0}>$ and $|\varphi_{0}>$ in
this case, the moments we now need to compute are
$<\phi_{0}|{\bf H}^{l}|\phi_{0}>$ and
$<\varphi_{0}|{\bf H}^{l}|\varphi_{0}>$. It is straightforward to
find that
\widetext
\begin{eqnarray}
<\phi_{0}|{\bf H}^{l}|\phi_{0}>&=&<1|{\bf H}^{l}|1>+2\,<1|{\bf
H}^{l}|3>+2\,<1|{\bf H}^{l}|5>  \nonumber \\
& & \mbox{}+{\cal P}\,(<1|{\bf H}^{l}|24>+2\,<1|{\bf H}^{l}|22>+2\,<1|{\bf
H}^{l}|19>),
\end{eqnarray}
and
\begin{eqnarray}
<\varphi_{0}|{\bf H}^{l}|\varphi_{0}>&= &<1|{\bf H}^{l}|1>-\,<1|{\bf
H}^{l}|2>-\,<1|{\bf H}^{l}|3>-\,<1|{\bf H}^{l}|7>+2\,<1|{\bf H}^{l}|4>+
\nonumber \\
& & \mbox{}+ {\cal P}\,(<1|{\bf H}^{l}|24>-\,<1|{\bf H}^{l}|23>-\,<1|{\bf
H}^{l}|22>  \nonumber \\
& & \mbox{}-\,<1|{\bf H}^{l}|21>+2\,<1|{\bf H}^{l}|20>).
\end{eqnarray}

\noindent
Notice that the site labels on the right hand sides of the above
two equations refer to Fig.~2. Through the techniques for
the calculation of these quantities previously discussed, the moments
$<\phi_{0}|{\bf H}^{l}|\phi_{0}>$ and
$<\varphi_{0}|{\bf H}^{l}|\varphi_{0}>$ can be readily obtained.

Anticipating the termination at $b_{4}^{2}$, we only need these
moments up to order $8$. In Table XII we give the moments with respect to
$|\phi_{0}>$ and $|\varphi_{0}>$. By utilizing
Eqs.~(14-16), we thus obtain $a_{n}$'s and $b_{n+1}^{2}$'s which
are consistent with those (in Table II) derived directly from
the recursion method.
To illustrate this consistency, we explicitly present here the results
for the $b_{n+1}^{2}$'s. From $<\phi_{0}|{\bf H}^{l}|\phi_{0}>$,
we have $b_{1}^{2}=t^{2}$, $b_{2}^{2}=2$, $b_{3}^{2}=1$, and
$b_{4}^{2}=(1-{\cal P}^{2})\,t^{2}=0$. From $<\varphi_{0}|{\bf
H}^{l}|\varphi_{0}>$,
we have $b_{1}^{2}=1$, $b_{2}^{2}=t^{2}$, $b_{3}^{2}=1$, and
$b_{4}^{2}=(1-{\cal P}^{2})\,t^{2}=0$. We thus demonstrate the fact that
the same results for the parameters $a_{n}$ and $b_{n+1}^{2}$ can
be obtained by using either the recursion or moment methods. The
advantage of the recursion method lies in that we can simultaneously
generate
the states and the parameters. However, it is sometimes difficult
to derive the states and parameters when the recursion method is
applied to some starting state, for example, a single carbon atom
state $|j>$, while the moment method provides standard procedures
to calculate the parameters after the moments are obtained.

\section{Local density of states}

In the recursion and moment methods,
the diagonal element of the Green function $(E-{\bf H})^{-1}$ can be
expressed as a continued fraction \cite{8,12}
\widetext
\begin{eqnarray}
G_{0}(E)& =& <f_{0}|\frac{1}{E-{\bf H}}|f_{0}>  \nonumber \\
    & = &{ 1 \over\displaystyle E-a_{0}- {\strut b_{1}^{2}
\over\displaystyle E-a_{1}-{\strut b_{2}^{2}
\over\displaystyle E-a_{2}-\cdots-{\strut b_{N-1}^{2} \over\displaystyle
E-a_{N-1}}}}}.
\end{eqnarray}

\noindent
The local density of states  $\rho(E)$ for $|f_{0}>$
is related to the imaginary part of $G_{0}(E)$ by
\begin{equation}
\rho(E)=\lim_{ \varepsilon \rightarrow 0}\,
-\frac{1}{\pi}\,{\rm Im}\ G_{0}(E+i \varepsilon).
\end{equation}
{}From the computational point of view,
$G_{0}(E)$ can be obtained by iteratively applying
the following transformation
$$
G_{n}(E)=\frac{1}{E-a_{n}-b_{n+1}^{2}G_{n+1}(E)}, $$
starting from $G_{N-1}(E)=1/(E-a_{N-1}).$
By plugging the parameters $a_{n}$ and $b_{n+1}^{2}$
from either one of the two methods into Eq.~(21) and using Eq.~(22),
we thus obtain the
local densities of states on several initial states.
In our two alternative applications of the recursion
method, we have used four different
starting states: $|u_{0}>$, $|v_{0}>$, $|\phi_{0}>$
and $|\varphi_{0}>$. Their $G_{0}(E)$'s can be written explicitly as
\widetext
\begin{equation}
G_{0}^{-1}=E+2-t^{2}\,\{E-2\,[E+t-(E+1+{\cal G}_{p})^{-1}]^{-1}\}^{-1}
\end{equation}
and
\begin{equation}
G_{0}^{-1}=E-t-1-\,\{E-t^{2}\,[E-1-(E-{\cal G}_{h})^{-1}]^{-1}\}^{-1},
\end{equation}
where
$${\cal G}_{p}=\left\{ \begin{array}{ll}
{\cal P}\,t & \mbox{for the state $|\phi_{0}>$} \\
-t^{2}\,\{E+1-\,[E+t-2\,(E-t^{2}\,(E+2)^{-1})^{-1}]^{-1}\}^{-1} &
\mbox{for the state $|u_{0}>$}
\end{array}
\right.  $$
and
$${\cal G}_{h}=\left\{ \begin{array}{ll}
{\cal P}\,t & \mbox{for the state $|\varphi_{0}>$} \\
t^{2}\,\{E-\,[E-1-t^{2}\,(E-\,(E-t-1)^{-1})^{-1}]^{-1}\}^{-1} &
\mbox{for the state $|v_{0}>$}.
\end{array}
\right.  $$

The LDOS around these ring-clusters are plotted in Fig.~3 and Fig.~4.
Notice that around a pentagon (hexagon) ring the LDOS is large at low
(high) energies. This is related to the fact that a pentagon (hexagon)
has zero (three) double bonds and five (three) single bonds. Also from
the path-integral approach, the local densities of states on two
antipode carbon atoms are plotted in Fig.~5. In principle, they are
experimentally accessible by using a scanning tunneling microscope
\cite{9}. In one of the path-integral method approaches, the initial
state is a $2p_{z}$ orbital on a carbon atom. The local density of
states in the case $t=1$ is plotted in Fig.~6. From Fig.~6, we can
obtain the degree of degeneracy for each energy level which is the
respective LDOS value times $2$. The common factor of $2$ comes from
the consideration that the total number of $\pi$ electrons is $60$. For
$t \neq 1$, conclusions for the degeneracy can be similarly drawn.

\section{Solution for the Electronic $\sigma$ states}
It is apparent that the Hamiltonian ${\bf H}_{\sigma}$ is more complex
than ${\bf H}_{\pi}$ due to the two different couplings. However, an
analytic transformation of this Hamiltonian into a simpler one with a
single ``renormalized" hopping parameter between sites can be
established \cite{10}. The energy states can then be readily solved.
The details of this transformation as well as the subsequent solutions
for the $\sigma$ states are given in Appendix C. As a result, we obtain
$90$ bonding $\sigma$ states and $90$ antibonding $\sigma$ states.
Among the bonding (antibonding) states, $30$ states are lumped together
at the energy level $V_{1}-V_{2}$ ($V_{1}+V_{2}$). The other $60$
$\sigma$ bonding and $60$ $\sigma$ antibonding states are closely
related to the energy spectrum for the $\pi$ states. To be more
specific, the eigenvalues of the other $60$ bonding (designated by plus
sign) and $60$ antibonding $\sigma$ states (designated by minus sign)
are
\begin{equation}
-\frac{V_{1}}{2} \pm \ V_{2}\ \sqrt{1+\frac{V_{1}}{V_{2}}\ E_{\lambda}+\frac{9\
V_{1}^{2}}{4\ V_{2}^{2}}},
\end{equation}
where $E_{\lambda}$ are the $\pi$ state eigenvalues listed in Table III.

\section{Conclusion}

In conclusion, we use several approaches based on the recursion and
path-integral methods in order to explore the electronic structure of a
$C_{60}$ molecule, obtaining exact closed-form expressions for the for
the $\pi$ and $\sigma$ eigenvalues and eigenfunctions, including the
HOMO and LUMO states, as well as the Green's functions and LDOS through
alternative methods. These quantities are relevant to the several
important experimental techniques which probe the local spectroscopy of
molecules (for instance, by using a scanning tunneling microscope, as
described in the review in Ref.~[8]). For comparison purposes, we have
also done a direct numerical diagonalization of the full Hamiltonian
and the results obtained are consistent with those from the previous
analytical methods.  However, the much more elegant and powerful
recursion and moment methods provide valuable insights and closed-form
expressions for various aspects of the electronic structure of
$C_{60}$.

\acknowledgments
The authors gratefully acknowledge stimulating and useful
discussions with C. N. Yang
and F. Guinea. We thank R. Richardson for his critical reading of the
manuscript. FN acknowledges partial support from a GE fellowship,
a Rackham grant, the NSF through grant DMR-90-01502, and SUN Microsystems.

\appendix

\section{Origin of $V_{1}$ and $V_{2}$}

In this appendix we examine more closely the physical interpretation for the
hopping integrals $V_{1}$ and $V_{2}$. Let $|i,\alpha>$
be a hybrid orbital located at site $i$. We have
\begin{equation}
|i,sp^{2}_{a}>=a_{s}^{i}\,|2s>+a_{x}^{i}\,|2p_{x}>+a_{y}^{i}\,|2p_{y}>,
\end{equation}
\begin{equation}
|i,sp^{2}_{b}>=b_{s}^{i}\,|2s>+b_{x}^{i}\,|2p_{x}>+b_{y}^{i}\,|2p_{y}>,
\end{equation}
and
\begin{equation}
|j,sp^{2}_{a}>=a_{s}^{j}\,|2s>+a_{x}^{j}\,|2p_{x}>+a_{y}^{j}\,|2p_{y}>,
\end{equation}
where $i$ and $j$ are nearest neighbors. We also assume that
orbitals $|i,sp^{2}_{a}>$ and $|j,sp^{2}_{a}>$ lie along the
bond connecting $i$ and $j$. It is straightforward to obtain
\widetext
\begin{equation}
-V_{1}=<i,sp^{2}_{a}|{\bf H}_{\sigma}|i,sp^{2}_{b}>
=a_{s}^{i\,\ast}b_{s}^{i}\,\epsilon_{s}+(a_{x}^{i\,\ast}b_{x}^{i}+a_{y}^{i\,\ast}b_{y}^{i})\,\epsilon_{p},
\end{equation}
and
\begin{eqnarray}
-V_{2}& =& <i,sp^{2}_{a}|{\bf H}_{\sigma}|j,sp^{2}_{a}> \nonumber \\
&=&a_{s}^{i\,\ast}a_{s}^{j}\,t_{ss}+a_{x}^{i\,\ast}a_{x}^{j}\,t_{p_{x}p_{x}}+a_{y}^{i\,\ast}a_{y}^{j}\,t_{p_{y}p_{y}} \nonumber \\
& & \mbox{}+(a_{s}^{i\,\ast}a_{x}^{j}+a_{x}^{i\,\ast}
a_{s}^{j})\,t_{sp_{x}}+(a_{s}^{i\,\ast}a_{y}^{j}+a_{y}^{i\,\ast}a_{s}^{j})\,t_{sp_{y}}+(a_{x}^{i\,\ast}a_{y}^{j}+a_{y}^{i\,\ast}a_{x}^{j})\,t_{p_{x}p_{y}}.
\end{eqnarray}

\noindent
Here $\epsilon_{s}$ and $\epsilon_{p}$ are the $2s$-level and $2p$-level
energies and the $t$'s are hopping matrix elements between orbitals on
nearest-neighboring sites. It is evident that, with a suitable choice of a
local coordinate system, we can always have
\widetext
\begin{equation}
<i,sp^{2}_{a}|{\bf H}_{\sigma}|i,sp^{2}_{b}>=<i,sp^{2}_{a}|{\bf
H}_{\sigma}|i,sp^{2}_{c}>=<i,sp^{2}_{b}|{\bf H}_{\sigma}|i,sp^{2}_{c}>
\end{equation}
and
\begin{equation}
<i,sp^{2}_{a}|{\bf H}_{\sigma}|j,sp^{2}_{a}> =<i,sp^{2}_{b}|{\bf
H}_{\sigma}|k,sp^{2}_{c}> =<i,sp^{2}_{c}|{\bf H}_{\sigma}|l,sp^{2}_{b}>.
\end{equation}

\noindent
In the above equation, we have assumed that orbitals $|i,sp^{2}_{b}>$,
$|k,sp^{2}_{c}>$ lie along the bond connecting $i$ and $k$ and orbitals
$|i,sp^{2}_{c}>$, $|l,sp^{2}_{b}>$ along the bond connecting $i$ and $l$.

\section{Calculation of ${\bf H}\,|\phi_{3}>$ and ${\bf H}\,|\varphi_{3}>$}
In this appendix we present the calculation of the recurrence relations
for states $|\phi_{3}>$ and $|\varphi_{3}>$ in the recursion method
case ${\cal B}$. It is shown
that the recurrence terminates at these states and two solutions for
$a_{3}$ are constructed from each state.
\widetext
\begin{eqnarray}
{\bf H}\,|\phi_{3}>&  = & \frac{1}{\sqrt{20}}\ \sum_{j=11}^{20}\ (|j>+ {\cal P}
|j'>)+\frac{1}{\sqrt{20}}\ [\,\sum_{j=21}^{30}\ (|j>+t\,|j'>)+{\cal P}\
\sum_{j=21}^{30}\ (t\,|j>+|j'>)] \nonumber  \\
&  =& |\phi_{2}>-\{-\frac{1}{\sqrt{20}}\ \sum_{j=21}^{30}\ [(1+{\cal
P}\,t)\,|j>+(t+{\cal P})\,|j'>]\}    \nonumber \\
&  =& |\phi_{2}>-\{-\frac{1}{\sqrt{20}}\ \sum_{j=21}^{30}\ [(1+{\cal
P}\,t)\,|j>+(1+{\cal P}\,t)\,{\cal P}\,|j'>]\}    \nonumber \\
&  =& |\phi_{2}>-(1+{\cal P}\,t)\ |\phi_{3}>.
\end{eqnarray}

\begin{eqnarray}
{\bf H}\,|\varphi_{3}>&  =& |\varphi_{2}>
+\frac{1}{\sqrt{12}}\,[t\,(|23'>-|24'>+|26'>-|27'>+|29'>-|30'>) \nonumber \\
 &   & \mbox{}+{\cal P}\,t\,(|23>-|24>+|26>-|27>+|29>-|30>)] \nonumber \\
&  =& |\varphi_{2}>+ {\cal
P}\,t\,\frac{1}{\sqrt{12}}\,[\,(|23>-|24>+|26>-|27>+|29>-|30>) \nonumber \\
&   & \mbox{}+{\cal P}\,(|23'>-|24'>+|26'>-|27'>+|29'>-|30'>)\,] \nonumber \\
&  =& |\varphi_{2}>+ {\cal P}\,t\ |\varphi_{3}>.
\end{eqnarray}

\section{One band-two band transformation}
In this appendix we derive the transformation of the $\sigma$-states
Hamiltonian into a $\pi$-type one. The resulting solutions of the eigenvalues
for the  $\sigma$ states are also presented. We first write the $\sigma$ states
eigenfunctions as
\begin{equation}
|\Psi_{\sigma}>=\sum_{i}\ {\cal C}_{i}\ |i,\alpha>,
\end{equation}
where $i$ runs over the atoms and ${\cal C}=a, b$ and $c$ represents the
corresponding coefficient for the orbital $\alpha$ at each site. Let us now
focus attention on a given atom labelled $i$. Also let atoms $j, k$ and $l$ be
the three adjacent nearest neighbors of $i$-atom. The energy eigenvalue
equation ${\bf H}_{\sigma}\ |\Psi_{\sigma}>=\epsilon \ |\Psi_{\sigma}>$ then
reads:
\begin{equation}
\epsilon \ a_{i}=-V_{2}\ a_{j}-V_{1}\ (b_{i}+c_{i}),
\end{equation}
\begin{equation}
\epsilon \ b_{i}=-V_{2}\ c_{k}-V_{1}\ (a_{i}+c_{i}),
\end{equation}
\begin{equation}
\epsilon \ c_{i}=-V_{2}\ b_{l}-V_{1}\ (a_{i}+b_{i}),
\end{equation}
and
\begin{equation}
\epsilon \ a_{j}=-V_{2}\ a_{i}-V_{1}\ (b_{j}+c_{j}),
\end{equation}
\begin{equation}
\epsilon \ c_{k}=-V_{2}\ b_{i}-V_{1}\ (a_{k}+b_{k}),
\end{equation}
\begin{equation}
\epsilon \ b_{l}=-V_{2}\ c_{i}-V_{1}\ (a_{l}+c_{l}),
\end{equation}
and so on. We then define
\begin{equation}
A_{i}=a_{i}+b_{i}+c_{i},
\end{equation}
as the sum of the coefficients for orbitals at the site $i$, and
\begin{equation}
B_{i}=a_{j}+c_{k}+b_{l},
\end{equation}
as the sum of the coefficients for orbitals on the bonds associated with the
site $i$. Similar equations hold for the other $A$'s and $B$'s.
In the first place, Eqs.~(C2) and (C5) can be rewritten as
\begin{equation}
(\epsilon - V_{1})\ a_{i}+V_{2}\ a_{j}=-V_{1}\ A_{i},
\end{equation}
and
\begin{equation}
V_{2}\ a_{i}+(\epsilon -V_{1})\ a_{j}=-V_{1}\ A_{j}.
\end{equation}
When the situation $A_{i}=A_{j}=0$ occurs, the secular equation immediately
yields
\begin{equation}
\epsilon =V_{1} \pm V_{2}.
\end{equation}
In all, there are thirty pairs of similar equation sets. We thus have two
energy levels $V_{1} - V_{2}$ belonging to the bonding states and $V_{1} +
V_{2}$ belonging to the antibonding states, each having a degree of degeneracy
equal to $30$.
Secondly, by respectively summing up Eqs.~(C2) through (C4) and Eqs.~(C5)
through (C7), we have
 \begin{equation}
\epsilon \ A_{i}=-V_{2}\ B_{i}-2\ V_{1}\ A_{i},
\end{equation}
and
\begin{equation}
(\epsilon -V_{1})\ B_{i}=-V_{2}\ A_{i}-V_{1}\ (A_{j}+A_{k}+A_{l}).
\end{equation}
By substituting Eq.~(C13) into Eq.~(C14), we obtain
\begin{equation}
[(\epsilon +2\ V_{1})(\epsilon -\ V_{1})-V_{2}^{2}]\ A_{i}= V_{1}\ V_{2}\
(A_{j}+A_{k}+A_{l}).
\end{equation}
It can be directly recognized that Eq.~(C15) is entirely equivalent to the
problem of a tight-binding Hamiltonian with one state per atom and a single
nearest-neighbor hopping integral, whose eigenvalue and eigenvector solutions
are already fully explored in the text. Let $E_{\lambda}$ ($\lambda
=1,2,\cdots,15$) stand for the eigenvalues summarized in Table III. It follows
then that
\begin{equation}
(\epsilon_{\lambda} +2\ V_{1})(\epsilon_{\lambda} -\ V_{1})-V_{2}^{2}=V_{1}\
V_{2}\ E_{\lambda}.
\end{equation}
Thus, we obtain
\begin{eqnarray}
\epsilon_{\lambda}& =&-\frac{V_{1}}{2} \pm
\ \sqrt{\frac{V_{1}^{2}}{4}+2\ V_{1}^{2}+V_{2}^{2}+V_{1}\ V_{2}\ E_{\lambda}}
\nonumber \\
& =&-\frac{V_{1}}{2} \pm
\ V_{2}\ \sqrt{1+\frac{V_{1}}{V_{2}}\ E_{\lambda}+\frac{9\ V_{1}^{2}}{4\
V_{2}^{2}}}.
\end{eqnarray}
The minus sign designates the bonding states, and the plus sign designates the
antibonding states. Finally, in the limit $V_{2} \gg V_{1}$, we have
\begin{equation}
\epsilon_{\lambda} \simeq -\frac{V_{1}}{2} \pm
\ V_{2}\ \pm \frac{V_{1}}{2}\ E_{\lambda}.
\end{equation}
It is also straightward to obtain the corresponding eigenvector for an
eigenvalue $\epsilon_{\lambda}$. Since all the $A_{i}$'s are already known, the
  initial coefficients $a_{i}, b_{i}$ and $c_{i}$ can be calculated.

\newpage
\begin{table}
\caption{States and parameters for case $\cal A$ in the recursion method
approach. It is interesting to notice that $a_{n}=a_{7-n}$ and $b_{n}=b_{8-n}$
for $n=4, 5, 6, 7$.}
\begin{tabular}{lccccccc}
\  &\multicolumn{3}{c}{Starting from a pentagon ring}
&\multicolumn{3}{c}{Starting from a hexagon ring } \\
$n$ & $|u_{n}>$ & $a_{n}$  & $b_{n}$ & $\ $ & $|v_{n}>$ & $a_{n}$  & $b_{n}$ \\
\hline
$0$ & $\frac{1}{\sqrt{5}}\,\sum_{j=1}^{5}\ |j>$ & $-2$ & $\ $ & $\ $ &
$\frac{1}{\sqrt{6}}\,\sum_{j=1}^{6}\ (-1)^{j+1}|j>$ & $1+t$ & $\ $    \\

$1$ & $-\frac{1}{\sqrt{5}}\,\sum_{j=6}^{10}\ |j>$ & $0$ & $t$ & $\ $ &
$\frac{1}{\sqrt{6}}\,\sum_{j=7}^{12}\ (-1)^{j}|j>$ & $0$ & $1$ \\

$2$ & $\frac{1}{\sqrt{10}}\,\sum_{j=11}^{20}\ |j>$ & $-t$ & $\sqrt{2}$  & $\ $
&  $\frac{1}{\sqrt{6}}\,\sum_{j=14}^{15}\ (-1)^{j+1}[|j>+|j+3>+|j+6>]$ & $1$ &
$t$ \\

$3$ & $-\frac{1}{\sqrt{10}}\,\sum_{j=21}^{30}\ |j>$ & $-1$ & $1$  & $\ $ &
$\frac{1}{\sqrt{6}}\,\sum_{j=23}^{24}\ (-1)^{j+1}[|j>+|j+3>+|j+6>]$ & $0$ & $1$
\\

$4$ & $\frac{1}{\sqrt{10}}\,\sum_{j=21}^{30}\ |j'>$ & $-1$ & $t$  & $\ $ &
$\frac{1}{\sqrt{6}}\,\sum_{j=23}^{24}\ (-1)^{j+1}[|j'>+|(j+3)'>+|(j+6)'>]$ &
$0$ & $t$ \\

$5$ & $-\frac{1}{\sqrt{10}}\,\sum_{j=11}^{20}\ |j'>$ & $-t$ & $1$ & $\ $ &
$\frac{1}{\sqrt{6}}\,\sum_{j=14}^{15}\ (-1)^{j+1}[|j'>+|(j+3)'>+|(j+6)'>]$ &
$1$ & $1$ \\

$6$ & $\frac{1}{\sqrt{5}}\,\sum_{j=6}^{10}\ |j'>$ & $0$ & $\sqrt{2}$  & $\ $ &
$\frac{1}{\sqrt{6}}\,\sum_{j=7}^{12}\ (-1)^{j}|j'>$ & $0$ & $t $ \\

$7$ & $-\frac{1}{\sqrt{5}}\,\sum_{j=1}^{5}\ |j'>$ & $-2$ & $t$ &  $\ $ &
$\frac{1}{\sqrt{6}}\,\sum_{j=1}^{6}\ (-1)^{j+1}|j'>$ & $1+t$ & $1 $ \\
\end{tabular}
\label{table1}
\end{table}
\begin{table}
\caption{States and parameters for case $\cal B$ in the recursion method
approach. The parity $\cal P$ can be $+1$ or $-1$.}
\begin{tabular}{lccc}
\  &\multicolumn{3}{c}{Starting from two opposite pentagon rings}
  \\
$n$ & $|\phi_{n}>$ & $a_{n}$  & $b_{n}$ \\
\hline

$0$ & $\frac{1}{\sqrt{10}}\,\sum_{j=1}^{5}\ [|j>+ {\cal P}\,|j'>]$ & $-2$  & $\
$     \\

$1$ & $-\frac{1}{\sqrt{10}}\,\sum_{j=6}^{10}\ [|j>+ {\cal P}\,|j'>]$ & $0$ &
$t$   \\

$2$ & $\frac{1}{\sqrt{20}}\,\sum_{j=11}^{20}\ [|j>+ {\cal P}\,|j'>]$ & $-t$ &
$\sqrt{2}$ \\

$3$ & $-\frac{1}{\sqrt{20}}\,\sum_{j=21}^{30}\ [|j>+ {\cal P}\,|j'>]$ &
$-(1+{\cal P}\,t)$ & $1$   \\ \hline\hline

\  &\multicolumn{3}{c}{Starting from two opposite hexagon rings} \\

$n$ & $|\varphi_{n}>$ & $a_{n}$  & $b_{n}$  \\
\hline

$0$ & $\frac{1}{\sqrt{12}}\,\sum_{j=1}^{6}\ (-1)^{j+1}[|j>+{\cal P}\,|j'>]$ &
$1+t$ & $\ $    \\

$1$ & $\frac{1}{\sqrt{12}}\,\sum_{j=}^{12}\ (-1)^{j}[|j>+{\cal P}\,|j'>]$ & $0$
& $1$ \\

$2$ &  $\frac{1}{\sqrt{12}}\,\sum_{j=14}^{15}\ (-1)^{j+1}[|j>+|j+3>+|j+6>+{\cal
P}\,(|j'>+|(j+3)'>+|(j+6)'>)]$ & $1$ & $t$ \\

$3$ &  $\frac{1}{\sqrt{12}}\,\sum_{j=23}^{24}\ (-1)^{j+1}[|j>+|j+3>+|j+6>+{\cal
P}\,(|j'>+|(j+3)'>+|(j+6)'>)]$ & ${\cal P}\,t$ & $1$ \\
\end{tabular}
\label{table2}
\end{table}

\begin{table}
\caption{The eigenvalues $E_{\lambda}$ and the corresponding eigenvectors
$\frac{1}{{\cal N}_{\lambda}}\,\sum_{n=0}^{3} P_{n}(E_{\lambda})|f_{n}>$ for
$C_{60}$ with $t=1$. Here $|\phi_{n}^{\pm}>$ ($|\varphi_{n}^{\pm}>$) denotes
$|\phi_{n}>$ ($|\varphi_{n}>$) with ${\cal  P}=\pm 1$.}

\begin{tabular}{cccccc}

$E_{\lambda}$ & $P_{1}(E_{\lambda})$ &  $P_{2}(E_{\lambda})$ &
 $P_{3}(E_{\lambda})$ & ${\cal N}_{\lambda}$ & $|f_{n}>$\\ \hline

$-3$ & $-1$ & $\sqrt{2}$ & $-\sqrt{2}$ & $6$ & $|\phi_{n}^{+}>$\\

$-1$ & $1$ &  $-\sqrt{2}$ & $-\sqrt{2}$ & $6$ & $|\phi_{n}^{+}>$\\

$(-1+\sqrt{13})/2$ & $(3+\sqrt{13})/2$ & $\sqrt{2}(3+\sqrt{13})/4$ &
$\sqrt{2}/2$ & $(39+9\sqrt{13})/4$ & $|\phi_{n}^{+}>$\\
$(-1-\sqrt{13})/2$ & $(3-\sqrt{13})/2$ & $\sqrt{2}(3-\sqrt{13})/4$ &
$\sqrt{2}/2$ & $(39-9\sqrt{13})/4$ & $|\phi_{n}^{+}>$\\
$\ $ & $\ $ & $\ $ & $\ $ & $\ $ &\ \\

$(-3+\sqrt{5}+\alpha)/4$ & $\frac{5+\sqrt{5}+\alpha}{4}$ &
$\frac{\sqrt{2}\,[6+2\sqrt{5}+(\sqrt{5}+1)\,\alpha]}{16}$ &
$\frac{\sqrt{2}\,[-4+8\sqrt{5}+(\sqrt{5}-1)\,\alpha]}{16}$ &
$\frac{95+5\sqrt{5}}{8}+\frac{(25+\sqrt{5})\,\alpha}{16}$ & $|\phi_{n}^{-}>$\\

$(-3+\sqrt{5}-\alpha)/4$ & $\frac{5+\sqrt{5}-\alpha}{4}$ &
$\frac{\sqrt{2}\,[6+2\sqrt{5}-(\sqrt{5}+1)\,\alpha]}{16}$ &
$\frac{\sqrt{2}\,[-4+8\sqrt{5}-(\sqrt{5}-1)\,\alpha]}{16}$ &
$\frac{95+5\sqrt{5}}{8}-\frac{(25+\sqrt{5})\,\alpha}{16}$ & $|\phi_{n}^{-}>$\\

$(-3-\sqrt{5}+\beta)/4$ & $\frac{5-\sqrt{5}+\beta}{4}$ &
$\frac{\sqrt{2}\,[6-2\sqrt{5}-(\sqrt{5}-1)\,\beta]}{16}$ &
$\frac{\sqrt{2}\,[-4-8\sqrt{5}-(\sqrt{5}+1)\,\beta]}{16}$ &
$\frac{95-5\sqrt{5}}{8}+\frac{(25-\sqrt{5})\,\beta}{16}$ & $|\phi_{n}^{-}>$\\

$-3-\sqrt{5}-\beta)/4$ & $\frac{5-\sqrt{5}-\beta}{4}$ &
$\frac{\sqrt{2}\,[6-2\sqrt{5}+(\sqrt{5}-1)\beta]}{16}$ &
$\frac{\sqrt{2}\,[-4-8\sqrt{5}+(\sqrt{5}+1)\beta]}{16}$ &
$\frac{95-5\sqrt{5}}{8}-\frac{(25-\sqrt{5})\,\beta}{16}$ & $|\phi_{n}^{-}>$\\
$\ $ & $\ $ & $\ $ & $\ $ & $\ $ & \ \\

$2$ & $0$ & $-1$ & $-1$ & $3$ & $|\varphi_{n}^{+}>$\\

$-1$ & $-3$ &  $2$ & $-1$ & $15$ & $|\varphi_{n}^{+}>$\\

$(3+\sqrt{5})/2$ & $(-1+\sqrt{5})/2$ & $(-1+\sqrt{5})/2$ & $(3-\sqrt{5})/2$ &
$(15-5\sqrt{5})/2$ & $|\varphi_{n}^{+}>$\\

$(3-\sqrt{5})/2$ & $(-1-\sqrt{5}/2$ & $(-1-\sqrt{5})/2$ & $(3+\sqrt{5})/2$ &
$(15+5\sqrt{5})/2$ & $|\varphi_{n}^{+}>$\\
$\ $ & $\ $ & $\ $ & $\ $ & $\ $ & \ \\

$(1+\sqrt{5})/2$ & $(-3+\sqrt{5})/2$ & $(-1-\sqrt{5})/2$ & $(1-\sqrt{5})/2$ &
$(15-3\sqrt{5})/2$ & $|\varphi_{n}^{-}>$\\

$(1-\sqrt{5})/2$ & $(-3-\sqrt{5})/2$ & $(-1+\sqrt{5})/2$ & $(1+\sqrt{5})/2$ &
$(15+3\sqrt{5})/2$ & $|\varphi_{n}^{-}>$\\

$(1+\sqrt{17})/2$ & $(-3+\sqrt{17})/2$ & $(5-\sqrt{17})/2$ & $-4+\sqrt{17}$ &
$51-12\sqrt{17}$ & $|\varphi_{n}^{-}>$\\

$(1-\sqrt{17})/2$ & $(-3-\sqrt{17})/2$ & $(5+\sqrt{17})/2$ & $-4-\sqrt{17}$ &
$51+12\sqrt{17}$ & $|\varphi_{n}^{-}>$\\
\end{tabular}
\label{table3}
\tablenotetext{Here $\alpha=\sqrt{2(19+\sqrt{5})}$ and
$\beta=\sqrt{2(19-\sqrt{5})}$.}
\end{table}

\narrowtext
\begin{table}
\caption{The eigenvalues and the corresponding degree of degeneracy
for $C_{60}$ with a single-bond hopping-integral $t_{1}$ and a double-bond
hopping-integral $t_{2}$. The characteristic polynomials from which those
eigenvalues are solved are indicated in the left-hand column.}
\begin{tabular}{ccc}
\ & Energy  & Degeneracy \\ \hline
\bigskip
 $P_{4}^{p+}$ & $\begin{array}{c}
-(2\,t_{1}+t_{2}) \\
 -(t_{1}+t_{2})\,/3+2\,\eta  \\ -(t_{1}+t_{2})\,/3-\eta+\sqrt{3}\,\xi \\
-(t_{1}+t_{2})\,/3-\eta-\sqrt{3}\,\xi \end{array}$
&  $\begin{array}{c}  1\\ 5 \\ 5 \\ 5 \end{array}$ \\
\bigskip
$P_{4}^{p-}$ & $\begin{array}{c} [(-3+\sqrt{5})\,t_{1}+\tau]\,/4 \\

 [(-3+\sqrt{5})\,t_{1}-\tau]\,/4 \\

 [(-3-\sqrt{5})\,t_{1}+\gamma]\,/4 \\

[(-3-\sqrt{5})\,t_{1}-\gamma]\,/4  \end{array}$
& $\begin{array}{c}  3\\ 3 \\ 3 \\ 3 \end{array}$ \\
\bigskip
$P_{4}^{h+}$ & $\begin{array}{c} (t_{1}+\sqrt{5\,t_{1}^{2}+4\,t_{2}^{2}})\,/2
\\ (t_{1}-\sqrt{5\,t_{1}^{2}+4\,t_{2}^{2}})\,/2
\\ t_{2}+(1+\sqrt{5})\,t_{1}\,/2 \\
 t_{2}+(1-\sqrt{5})\,t_{1}\,/2  \end{array}$
& $\begin{array}{c}  4\\ 4 \\ 3 \\ 3  \end{array}$ \\

$P_{4}^{h-}$ &   $\begin{array}{c}
(t_{1}+\sqrt{5\,t_{1}^{2}-4\,t_{1}\,t_{2}+4\,t_{2}^{2}})\,/2 \\
(t_{1}-\sqrt{5\,t_{1}^{2}-4\,t_{1}\,t_{2}+4\,t_{2}^{2}})\,/2 \\
(t_{1}+\sqrt{5\,t_{1}^{2}+8\,t_{1}\,t_{2}+4\,t_{2}^{2}})\,/2 \\
(t_{1}-\sqrt{5\,t_{1}^{2}+8\,t_{1}\,t_{2}+4\,t_{2}^{2}})\,/2  \end{array}$
& $\begin{array}{c}  5\\ 5 \\ 4 \\ 4 \end{array}$ \\
\end{tabular}
\label{table4}
\tablenotetext{Here
$\tau=\sqrt{2\,[\,8\,t_{2}^{2}-(4+4\sqrt{5})\,t_{2}\,t_{1}+(15+5\sqrt{5})\,t_{1}^{2}]}$ and $\gamma=\sqrt{2\,[\,8\,t_{2}^{2}-(4-4\sqrt{5})\,t_{2}\,t_{1}+(15-5\sqrt{5})\,t_{1}^{2}]}$. Also $\eta$ and $\xi$ satisfy
$(16\,t_{2}^{3}-24\,t_{2}^{2}t_{1}+12\,t_{2}\,t_{1}^{2}+25\,t_{1}^{3})\,/54=\eta^{3}-3\eta \xi^{2}$
and
$\sqrt{3\,(64\,t_{2}^{4}-160\,t_{2}^{3}\,t_{1}+288\,t_{2}^{2}\,t_{1}^{2}-200\,t_{2}\,t_{1}^{3}+125\,t_{1}^{4})}\,/18=3\eta^{2}\xi-\xi^{3}$.}
\end{table}

\widetext
\begin{table}
\caption{The HOMO wave function $\frac{1}{{\cal N}_{\lambda}}\,\sum_{n=0}^{3}
P_{n}(E_{\lambda})|\varphi_{n}^{-}>$ and LUMO wave function $\frac{1}{{\cal
N}_{\lambda}}\,\sum_{n=0}^{3} P_{n}(E_{\lambda})|\phi_{n}^{-}>$ for $C_{60}$ in
terms of $t$. Here $|\varphi_{n}^{-}>$  stands for $|\varphi_{n}>$ with ${\cal
P}=-1$ and $|\phi_{n}^{-}>$ for $|\phi_{n}>$ with ${\cal P}=-1$.}
\begin{tabular}{ccc}
\bigskip
HOMO & $\begin{array}{c}
E_{\lambda} \\ P_{1}(E_{\lambda})  \\P_{2}(E_{\lambda}) \\ P_{3}(E_{\lambda})\\
{\cal N}_{\lambda} \end{array}$ &
$\begin{array}{c} (1-\sqrt{4\,t^{2}-4\,t+5})/\,2 \\
-t-(1+\sqrt{4\,t^{2}-4\,t+5})/\,2 \\
   t-(3-\sqrt{4\,t^{2}-4\,t+5})/\,2 \\
   t-(1-\sqrt{4\,t^{2}-4\,t+5})/\,2 \\
6\,t^{2}-6\,t-\frac{15}{2}+(3\,t-\frac{3}{2})\,\sqrt{4\,t^{2}-4\,t+5}
\end{array}$ \\
LUMO &  $\begin{array}{c}
E_{\lambda} \\ P_{1}(E_{\lambda})  \\P_{2}(E_{\lambda}) \\ P_{3}(E_{\lambda})\\
{\cal N}_{\lambda} \end{array}$ &
$\begin{array}{c}
(-3-\sqrt{5}+\gamma)\,/4  \\ (5-\sqrt{5}+\gamma)/4\,t \\
\sqrt{2}\,[\,10-6\,\sqrt{5}-4\,(1-\sqrt{5})\,t+(1-\sqrt{5})\,\gamma]/\,16\,t \\
-\sqrt{2}\,[\,4\,\sqrt{5}+4\,(1+\sqrt{5})\,t+(1+\sqrt{5})\gamma]/\,16\,t \\

 [\,150-50\,\sqrt{5}+80\,t^{2}-
40\,(1-\sqrt{5})\,t+(25-5\,\sqrt{5}+4\,\sqrt{5}\,t)\,\gamma]/\,16\,t^{2}
\end{array}$ \\
\end{tabular}
\label{table5}
\tablenotetext{Here
$\gamma=\sqrt{2\,[8\,t^{2}-4\,(1-\sqrt{5})\,t+15-5\sqrt{5}]}.$}
\end{table}

\narrowtext
\begin{table}
\caption{Moments (sum over paths) ${\cal M}_{9}$
through ${\cal M}_{30}$ for $C_{60}$ with $t=1$.}

\begin{tabular}{cr}
Order$(l)$  & Moment  $({\cal M}_{l})$   \\     \hline
$9$  & $-306$ \\
$10$  & $4274$ \\
$11$  & $-3080$ \\
$12$  & $31227$ \\
$13$  & $-29718$ \\
$14$  & $234559$ \\
$15$  & $ -279100$ \\
$16$  & $ 1803375 $ \\
$17$  & $-2572542$ \\
$18$  & $14149891$ \\
$19$  & $-23398880 $ \\
$20$  & $113056535 $ \\
$21$  & $-210843318$ \\
$22$  & $918114387 $ \\
$23$  & $-1887655172$ \\
$24$  & $7564926707$ \\
$25$  & $ -16828070362$ \\
$26$  & $63140353799$ \\
$27$  & $-149626028160 $ \\
$28$  & $532999985631$ \\
$29$  & $-1328522904154$ \\
$30$  & $4543918293899$ \\
\end{tabular}
\label{table6}
\end{table}

\widetext
\begin{table}
\caption{Parameters $a_{n}$ and $b_{n}$ calculated from the moments   ${\cal
M}_{l}$ with $t=1$ in the moment method for $C_{60}$. Numbers in
denominator are relatively prime to those in numerator and all numbers in
square root are primes.}
\begin{tabular}{ccc}
$n$ & $a_{n}$ & $b_{n}$\\     \hline
$0$    & $0$  &        $\ $                  \\
$1$   & $0$  &         $\sqrt{3}$  \\
$2$   & $-\frac{1}{3}$  &         $\sqrt{2}$    \\
$3$  & $\frac{11}{69}$  &          $\frac{1}{3} \sqrt{23}$         \\
$4$  & $-\frac{6633}{10925}$  &         $\frac{5}{23} \sqrt{2\cdot19}$  \\
$5$   & $-\frac{ 1109069}{2724600}$  &
$\frac{6}{475}\sqrt{2\cdot23\cdot239}$  \\
$6$  & $-\frac{ 52107413}{684218760 }$  & $\frac{5}{5736}
\sqrt{5\cdot19\cdot23857}$ \\
$7$  & $-\frac{66333080317}{113465204135}$  & $\frac{12}{ 119285
} \sqrt{23\cdot239\cdot41357}$ \\
$8$  & $-\frac{23875175834189}{52324285077614}$   & $\frac{1}{2853633 }
\sqrt{3\cdot5 \cdot11 \cdot31\cdot 80657 \cdot23857}$ \\
$9$   & $\frac{603116478351886109 }{ 1730319460378457102}$  & $\frac{23
}{165024222} \sqrt{ 3\cdot7\cdot17\cdot41357\cdot11492779}$ \\

$10$  & $-\frac{21011937073656177526747}{591749269376944653900009}$   &
$\frac{3}{31455736123} \sqrt{3\cdot11\cdot31\cdot137\cdot80657\cdot15257198651

}$ \\
$11$  & $\frac{90586832655047641731260522470}{156471266768327126141170869441}$
&    $\frac{7}{18812125936683} \sqrt{2 \cdot17\cdot23\cdot41\cdot67\cdot
11492779\cdot432553690463}$ \\

$12$  &
$-\frac{13031823151655892649412664693192093}{26315692870966750872559328607407057
}$   & $\frac{3}{8317574913913027}
\sqrt{137\cdot15257198651\cdot3163866047896700891
}$ \\

$13$  &
$-\frac{284524265315564231563836760572796495}{698223314042680095913984617907235831}$  &  $\frac{ 551}{3163866047896700891} \sqrt{7\cdot13\cdot41\cdot67\cdot 4544717\cdot 57739553 \cdot   432553690463 }$ \\

$14$  & $\frac{290400229158644153}{ 220686749524952341}$  &
$\frac{98208}{220686749524952341} \sqrt{3\cdot5
\cdot13\cdot17\cdot89\cdot3163866047896700891}$ \\
\end{tabular}
\label{table7}
\end{table}

\widetext
\begin{table}
\caption{Moments (sum over paths) ${\cal M}_{9}$
through ${\cal M}_{16}$ for $C_{60}$ with an arbitrary $t$.}
\begin{tabular}{cc}
Order$(l)$ & Moment  $({\cal M}_{l})$ \\  \hline
$9$  & $-54\,t^{4}-18\,t^{3}-162\,t^{2}-72$ \\
$10$ & $t^{10}+50\,t^{8}+40\,t^{7}+500\,t^{6}+300\,t^{5}+1470\,t^{4}+380\,t^{3
}+1280\,t^{2}+254$  \\
$11$ &$-154\,t^{6}-132\,t^{5}-946\,t^{4}-286\,t^{3}-1232\,t^{2}-330$  \\
$12$ & $t^{12}+72\,t^{10}+80\,t^{9}+1110\,t^{8}+1080\,t^{7}+5640\,t^{6}+3384\,
t^{5}+10200\,t^{4}+2520\,t^{3}+6192\,t^{2}+948$  \\
\bigskip
$13$ & $-364\,t^{8}-546\,t^{7}-3848\,t^{6}-2886\,t^{5}-10088\,t^{4}-2782\,t^{3
}-7774\,t^{2}-1430$  \\
\bigskip
$14$ &  $\begin{array}{l}
t^{14}+98\,t^{12}+140\,t^{11}+2156\,t^{10}+2940\,t^{9}+16870\,t^{8}+
16464\,t^{7}+ \\ 52444\,t^{6}+30128\,t^{5}+65016\,t^{4}+15232\,t^{3}
+29456\,t^{2}+3614 \end{array}$ \\
\bigskip
$15$ &  $\begin{array}{l}
-756\,t^{10}-1680\,t^{9}-12390\,t^{8}-15750\,t^{7}-55330\,t^{6} \\ -36936
\,t^{5}-84690\,t^{4}-21430\,t^{3}
-44130\,t^{2}-6008 \end{array}$ \\

$16$ &  $\begin{array}{l}
t^{16}+128\,t^{14}+224\,t^{13}+3808\,t^{12}+6720\,t^{11}+42560\,t^{10}
+58240\,t^{9}+203576\,t^{8}+ \\ 189696\,t^{7}+431264\,t^{6}+234464\,t^{5}+
392720\,t^{4}+87168\,t^{3}+138816\,t^{2}+13990 \end{array}$ \\
\end{tabular}
\label{table8}
\end{table}

\narrowtext
\begin{table}
\caption{Moments (sum over paths) for states $|I_{+}>$ and $|I_{-}>$ with
$t=1$.}
\begin{tabular}{crr}
Order$(l)$  & $<I_{+}|{\bf H}^{l}|I_{+}>$  & $<I_{-}|{\bf H}^{l}|I_{-}>$  \\
 \hline
$9$  & $-312$   & $-300$    \\
$10$  & $4319$    & $ 4231$   \\
$11$  & $-3278$   & $-2882$    \\
$12$  & $32339$  & $30115$    \\
$13$  & $-33436$  & $-26000$   \\
$14$  & $252339$  & $216779$    \\
$15$  & $\  $   & $ -225080$   \\
$16$  & $\   $ & $1571823$  \\
\end{tabular}
\label{table9}
\end{table}

\begin{table}
\caption{Parameters $a_{n}$ and $b_{n}$ computed from the moments
$<I_{\pm}|{\bf H}^{l}|I_{\pm}>$ with $t=1$.}
\begin{tabular}{ccccc}
\  &\multicolumn{2}{c}{From $<I_{+}|{\bf H}^{l}|I_{+}>$}
&\multicolumn{2}{c}{From $<I_{-}|{\bf H}^{l}|I_{-}>$} \\
$n$ & $a_{n}$ & $b_{n}$ & $a_{n}$ & $b_{n}$\\     \hline
$0$    & $0$  &        $\ $   & $0$  &        $\ $               \\
$1$   & $0$  & $\sqrt{3}$ & $0$  &         $\sqrt{3}$ \\
$2$   & $-\frac{1}{3}$  & $\sqrt{2}$ & $-\frac{1}{3}$  &   $\sqrt{2}$   \\
$3$  & $\frac{11}{69}$  & $\frac{1}{3} \sqrt{23}$  & $\frac{11}{69}$  &
$\frac{1}{3} \sqrt{23}$       \\
$4$  & $-\frac{18027}{21850}$  &         $\frac{5}{23} \sqrt{2\cdot19}$  &
$-\frac{1701}{4370}$  &         $\frac{5}{23} \sqrt{2\cdot19}$ \\
$5$   & $-\frac{1964421}{36832450}$  &
$\frac{1}{950}\sqrt{3\cdot23\cdot137\cdot283}$  & $-\frac{11563}{44650}$  &
$\frac{1}{190}\sqrt{3\cdot5\cdot23\cdot47}$\\
$6$  & $\frac{40798}{38771}$  & $\frac{855}{38771}
\sqrt{2\cdot3\cdot13\cdot19}$ & $\frac{342588}{22258025}$  & $\frac{19}{1175}
\sqrt{2\cdot997}$\\
$7$ &$\  $ & $\ $ & $-\frac{3661}{18943}$  & $\frac{100}{18943}
\sqrt{17\cdot47\cdot89}$  \\
\end{tabular}
\label{table10}
\end{table}

\widetext
\begin{table}
\caption{Moments (sum over paths) $<I_{\pm}|{\bf H}^{l}|I_{\pm}>$
($l=9,\cdots,16$) for an arbitrary $t$.}
\begin{tabular}{cc}
Order$(l)$     &  Moment $(<I_{+}|{\bf H}^{l}|I_{+}>)$   \\     \hline
$9$  & $-54\,t^{4}-24\,t^{3}-162\,t^{2}-72$ \\
$10$ & $t^{10}+50\,t^{8}+40\,t^{7}+500\,t^{6}+304\,t^{5}+1490\,t^{4}+400\,t^{3
}+1280\,t^{2}+254$  \\
$11$ &$-154\,t^{6}-176\,t^{5}-990\,t^{4}-396\,t^{3}-1232\,t^{2}-330$  \\
$12$ & $t^{12}+72\,t^{10}+80\,t^{9}+1110\,t^{8}+1104\,t^{7}+5784\,t^{6}+3648\,
t^{5}+10560\,t^{4}+2840\,t^{3}+6192\,t^{2}+948$  \\
\bigskip
$13$ & $-364\,t^{8}-728\,t^{7}-4212\,t^{6}-4056\,t^{5}-10920\,t^{4}-3952\,t^{3
}-7774\,t^{2}-1430$  \\
\bigskip
$14$ &  $\begin{array}{l}
t^{14}+98\,t^{12}+140\,t^{11}+2156\,t^{10}+3024\,t^{9}+17458\,t^{8}+
18032\,t^{7}+ \\ 55944\,t^{6}+35000\,t^{5}+69020\,t^{4}+18396\,t^{3}
+29456\,t^{2}+3614 \end{array}$ \\
\bigskip
$15$ &  $\begin{array}{l}
-756\,t^{10}-2240\,t^{9}-14070\,t^{8}-22380\,t^{7}-64470\,t^{6} \\ -53736
\,t^{5}-94230\,t^{4}-31100\,t^{3}
-44130\,t^{2}-6008 \end{array}$ \\
$16$ &  $\begin{array}{l}
t^{16}+128\,t^{14}+224\,t^{13}+3808\,t^{12}+6944\,t^{11}+44352\,t^{10}
+64512\,t^{9}+222168\,t^{8}+ \\ 225056\,t^{7}+481696\,t^{6}+292960\,t^{5}+
428144\,t^{4}+112128\,t^{3}+138816\,t^{2}+13990 \end{array}$ \\ \hline\hline
Order$(l)$     &  Moment $(<I_{-}|{\bf H}^{l}|I_{-}>)$   \\  \hline
$9$  & $-54\,t^{4}-12\,t^{3}-162\,t^{2}-72$ \\
$10$ & $t^{10}+50\,t^{8}+40\,t^{7}+500\,t^{6}+296\,t^{5}+1450\,t^{4}+360\,t^{3
}+1280\,t^{2}+254$  \\
$11$ &$-154\,t^{6}-88\,t^{5}-902\,t^{4}-176\,t^{3}-1232\,t^{2}-330$  \\
$12$ & $t^{12}+72\,t^{10}+80\,t^{9}+1110\,t^{8}+1056\,t^{7}+5596\,t^{6}+3120\,
t^{5}+9840\,t^{4}+2200\,t^{3}+6192\,t^{2}+948$  \\
\bigskip
$13$ & $-364\,t^{8}-364\,t^{7}-3484\,t^{6}-1716\,t^{5}-9256\,t^{4}-1612\,t^{3
}-7774\,t^{2}-1430$  \\
\bigskip
$14$ &  $\begin{array}{l}
t^{14}+98\,t^{12}+140\,t^{11}+2156\,t^{10}+2856\,t^{9}+116282\,t^{8}+
14896\,t^{7}+ \\ 48944\,t^{6}+25256\,t^{5}+61012\,t^{4}+12068\,t^{3}
+29456\,t^{2}+3614 \end{array}$ \\
\bigskip
$15$ &  $\begin{array}{l}
-756\,t^{10}-1120\,t^{9}-10710\,t^{8}-9120\,t^{7}-46190\,t^{6} \\ -20136
\,t^{5}-75150\,t^{4}-11760\,t^{3}
-44130\,t^{2}-6008 \end{array}$ \\
$16$ &  $\begin{array}{l}
t^{16}+128\,t^{14}+224\,t^{13}+3808\,t^{12}+6496\,t^{11}+40768\,t^{10}
+51968\,t^{9}+184984\,t^{8}+ \\ 154336\,t^{7}+380832\,t^{6}+175968\,t^{5}+
357296\,t^{4}+62208\,t^{3}+138816\,t^{2}+13990 \end{array}$ \\
\end{tabular}
\label{table11}
\end{table}

\begin{table}
\caption{Moments (sum over paths) $<\phi_{0}|{\bf H}^{l}|\phi_{0}>$ and
$<\varphi_{0}|{\bf H}^{l}|\varphi_{0}>$ for $l=1,\cdots,8.$}
\begin{tabular}{cc}
Order$(l)$     &  Moment $(<\phi_{0}|{\bf H}^{l}|\phi_{0}>)$   \\     \hline
$1$  &$-2$  \\
$2$  &$t^{2}+4$    \\
$3$  &$-4\,t^{2}-8$         \\
$4$  &$t^{4}+14\,t^{2}+16$  \\
$5$  &$-6\,t^{4}-2\,t^{3}-40\,t^{2}-32$  \\
$6$  & $t^{6}+30\,t^{4}+8\,t^{3}+110\,t^{2}+64$ \\
$7$  & $-8\,t^{6}-6\,t^{5}-112\,t^{4}-(36+2{\cal P})\,t^{3}-282\,t^{2}-128$ \\
$8$  & $t^{8}+52\,t^{6}+32\,t^{5}+(396+4 {\cal P})\,t^{4}+(116+12{\cal
P})\,t^{3}+708\,t^{2}+256$ \\ \hline\hline
Order$(l)$     &  Moment  $(<\varphi_{0}|{\bf H}^{l}|\varphi_{0}>)$   \\
\hline
$1$  &$t+1$  \\
$2$  &$t^{2}+2\,t+2$    \\
$3$  &$t^{3}+3\,t^{2}+5\,t+3$         \\
$4$  &$t^{4}+4\,t^{3}+10\,t^{2}+10\,t+5$  \\
$5$  &$t^{5}+5\,t^{4}+16\,t^{3}+25\,t^{2}+20\,t+8$  \\
$6$  & $t^{6}+6\,t^{5}+24\,t^{4}+48\,t^{3}+60\,t^{2}+38\,t+13$ \\
$7$  & $t^{7}+7\,t^{6}+33\,t^{5}+84\,t^{4}+(133+{\cal
P})\,t^{3}+133\,t^{2}+71\,t+21$ \\
$8$  & $t^{8}+8\,t^{7}+44\,t^{6}+132\,t^{5}+(266+2 {\cal P})\,t^{4}+(336+4
{\cal P})\,t^{3}+284\,t^{2}+130\,t+34$ \\
\end{tabular}
\label{table12}
\end{table}

\newpage
\begin{figure}
\caption{Flattened $C_{60}$ molecule obtained by stretching (a) a
pentagon and (b) a hexagon, with the site labels used in cases
$\cal A$ and $\cal B$ in the recursion method.}
\label{fig1}\end{figure}

\begin{figure}
\caption{Site labels used in the path-integral method.
Independent vertices are those with $j$ running from $1$ to $24$.}
\label{fig2}\end{figure}

\begin{figure}
\caption{The local density of states around ring-clusters : (a)
a pentagon, (b) a hexagon. They are obtained by using the recursion method.
Notice that around a pentagon (hexagon) ring the LDOS is large at low (high)
energies. This is related to the fact that a pentagon (hexagon) has zero
(three) double bonds and five (three) single bonds.}
\label{fig3}\end{figure}

\begin{figure}
\caption{The local density of states around two opposite ({\it i.e.},
antipodes) pentagons with (a) even parity, (b) odd parity, and two two opposite
hexagons with (c) even parity, (d) odd parity. They are obtained by using the
recursion method. Notice that around two antipode pentagon (hexagon) rings the
LDOS is large at low (high) energies.}
\label{fig4}\end{figure}

\begin{figure}
\caption{The local density of states on two opposite carbon atoms (antipodes)
with (a) even parity, (b) odd parity. They are obtained by
using the moment method. Notice that there are $7$ even parity solutions in (a)
and $8$ different odd parity solutions in (b); for a total of $15$ different
energy levels (for $t=1$).}
\label{fig5}\end{figure}

\begin{figure}
\caption{The local density of states for a carbon atom obtained from the
moment method with the respective degree of degeneracy shown
above each peak.}
\label{fig6}\end{figure}


\begin{references}
\bibitem{1} E. Manousakis, Phys. Rev. B {\bf 44}, 10991 (1991); {\bf 48},
2024(E) (1993).
\bibitem{2} D. Tom\'{a}nek and M.A. Schluter, Phys. Rev. Lett. {\bf 67}, 2331
(1991);
{\bf 56}, 1055 (1986); Phys. Rev. B {\bf 36}, 1208 (1987); D. Tom\'anek {\it et
al\/},
Phys. Rev. B {\bf 39}, 5361 (1989).
\bibitem{3}Y. Deng and C.N. Yang, Phys. Lett. A {\bf 170}, 116 (1992).
\bibitem{4}R. Friedberg, T.D. Lee, and H.C. Ren, Phys. Rev. B {\bf 46}, 14150
(1992).
\bibitem{5}J. Gonz\'{a}lez, F. Guinea, and M.A.H. Vozmediano,
Phys. Rev. Lett. {\bf 69}, 172 (1992); Nuclear Phys., in press; F. Guinea, J.
Gonz\'{a}lez, and M. A. H. Vozmediano, Phys. Rev. B {\bf 47}, 16576 (1993).
\bibitem{6} S. Satpathy, Chem. Phys. Lett. {\bf 130}, 545 (1986); S. Satpathy
{\it et al},
Phys. Rev. B {\bf 46}, 1773 (1992).
\bibitem{7}V. Elser and R.C. Haddon, Phys. Rev. A {\bf 36}, 4579 (1987);
Nature
{\bf 325}, 792 (1987).
\bibitem{8}R. Haydock, in {\it Solid State Physics}, edited by H. Ehrenreich,
F. Seitz, and D. Turnbull (Academic, New York, 1980), Vol. 35.
\bibitem{9}J. H. Weaver, Accounts of Chemical Research {\bf 25}, 143 (1992);
and references therein.
\bibitem{10} D. Weaire and M. F. Thorpe, in {\it Computational Methods for
Large Molecules and Localized States}, edited by F. Herman, A. D. McLean and R.
K. Nesbet (Plenum, New York, 1973).
\bibitem{11}L. D. Landau and E. M. Lifschitz, {\em Statistical Physics}
(Pergamon, Oxford, 1968), 2nd ed., pp. 447-454.
\bibitem{12}G. Grosso and G. Pastori Parravicini, in {\it Memory Function
Approaches to Stochastic Problems in Condensed Matter}, edited by M. W. Evans,
P. Grigolin, and G. Pastori Parravicini, Advances in Chemical Physics Vol 62
(Wiley, New York, 1985).

\end{references}
\end{document}